\documentclass[aps,preprintnumbers,showpacs,nofootinbib,floatfix]{revtex4}
\usepackage[english]{babel}
\usepackage{graphicx}
\usepackage{subfigure}
\usepackage{dcolumn}
\usepackage{bm}
\usepackage{amsmath}
\newcommand{\newc}{\newcommand}
\newc{\ra}{\rightarrow}
\newc{\lra}{\leftrightarrow}
\newc{\beq}{\begin{equation}}
\newc{\eeq}{\end{equation}}
\newc{\barr}{\begin{eqnarray}}
\newc{\earr}{\end{eqnarray}}

\def\vbf{\mbox{\boldmath $\upsilon$}}
\def\barr{\begin{eqnarray}}
\def\earr{\end{eqnarray}}

 \def\vbf{\mbox{\boldmath $\upsilon$}}

\begin{document}
\newcommand{\Od}{{\cal O}}
\newcommand{\lsim}   {\mathrel{\mathop{\kern 0pt \rlap
  {\raise.2ex\hbox{$<$}}}
  \lower.9ex\hbox{\kern-.190em $\sim$}}}
\newcommand{\gsim}   {\mathrel{\mathop{\kern 0pt \rlap
  {\raise.2ex\hbox{$>$}}}
  \lower.9ex\hbox{\kern-.190em $\sim$}}}

\title{ Inelastic WIMP-nucleus scattering  to the first excited state in $^{125}$Te.}

\author{J.D. Vergados$^{1,*}$, F.T.  Avignone III$^{2}$, M. Kortelainen$^{3,4}$, P. Pirinen$^3$, P. C. Srivastava$^{5}$,  J. Suhonen$^3$ and A. W. Thomas$^{1}$}
\affiliation{$^1$ ARC Centre of Excellence in Particle Physics at the Terascale and Centre for the Subatomic Structure of Matter (CSSM), University of Adelaide, Adelaide SA 5005, Australia,
\footnote{Permanent address:Theoretical Physics,University of Ioannina, Ioannina, Gr 451 10, Greece.}}
\affiliation{$^{(2)}${\it University of South Carolina, Columbia, SC 29208, USA, }}
\affiliation{$^3$University of Jyvaskyla, Department of Physics, P.O. Box 35, FI-40014 , Finland}
\affiliation{$^4$Helsinki Institute of Physics, P.O. Box 64, FI00014 University of Helsinki, Finland}
\affiliation{$^5$Department of Physics, Indian Institute of Technology, Roorkee 247667, India}
\vspace{0.5cm}
\begin{abstract}
 The direct detection of dark matter constituents, in particular the weakly interacting massive particles (WIMPs), is 
 considered central to
particle physics and cosmology.  In this paper we study transitions to the excited states, possible in some nuclei, which have sufficiently low lying excited states. Examples considered previously were  the first excited states of $^{127}$I and $^{129}$Xe and $^{83}$Kr. Here we examine $^{125}$Te, which offers some advantages and is currently being considered as a target.
In all these cases the extra signature of the gamma rays following the de-excitation of these states has definite advantages over the purely nuclear recoil and, in principle, such a signature can 
be exploited experimentally. A brief discussion of the experimental feasibility is given in the context of the CUORE experiment.
\end{abstract}
\date{\today}
\maketitle
\section{Introduction}
At present there exists plenty of evidence of the existence of dark matter from cosmological observations, 
the combined MAXIMA-1 \cite{MAXIMA1,MAXIMA2,MAXIMA3}, BOOMERANG \cite{BOOMERANG1,BOOMERANG2},
DASI \cite{DASI}, COBE/DMR Cosmic Microwave Background (CMB)
observations \cite{COBE} 
\cite{SPERGEL},  
 the recent WMAP  \cite{WMAP06} and Planck \cite{PlanckCP15} data 
as well as the observed rotational curves in the galactic halos, see e.g. the review \cite{UK01}. It is, however, essential to directly
detect such matter in order to 
unravel its nature.\\
 At present there exist many such candidates, called Weakly Interacting Massive Particles (WIMPs), e.g.  the
LSP (Lightest Supersymmetric Particle) \cite{ref2a,ref2b,ref2c,ref2d,ELLROSZ,GomVer01,GomLazPa1,GomLazPa2,ELLFOR}, technibaryon \cite{Nussinov92,GKS06}, mirror matter\cite{FLV72,Foot11} and Kaluza-Klein models with universal extra dimensions\cite{ST02a,OikVerMou}. These models predict an interaction of dark matter with ordinary matter via the exchange of a scalar particle, which leads to a spin independent interaction (SI) or vector boson interaction, which leads to a spin dependent (SD) nucleon cross section.    Additional theoretical tools are the structure of the nucleus, see e.g.  \cite{JDV06a,Dree00,Dree,Chen}, and the nuclear matrix elements \cite{Ress,DIVA00,JDV03,JDV04,VF07}. 

This paper will focus on the spin dependent WIMP nucleus interaction. This cross section can be sizable in a variety of models, including the lightest supersymmetric particle (LSP)  \cite{CHATTO,CCN03,JDV03, WELLS}, in the co-annihilation region \cite{Cannoni11}, where the ratio of the SD  to to the SI nucleon cross section, depending on $\tan{\beta}$ and the WIMP mass,  can be  large, e.g.  $10^3$ in the WIMP mass range 200-500 GeV. 
Furthermore more recent calculations in the supersymmetric $SO(10)$ model \cite{Gogoladze13}, also in the co-annihilation region, predict ratios
of the order of $2\times 10^{3}$  for a WIMP mass of about 850 GeV. Models of exotic WIMPs, like Kaluza-Klein models \cite{ST02a,OikVerMou} and Majorana particles with  spin  3/2 \cite{SavVer13},  can also lead to large nucleon spin induced cross sections, which satisfy the relic abundance constraint.  
This interaction is very important because it can lead to inelastic WIMP-nucleus scattering with a non negligible probability, provided that the energy of the excited state  is sufficiently low, 
a prospect proposed a long time
ago \cite{GOODWIT} and considered in some detail by Ejiri and collaborators \cite{EFO93}.
 For sufficiently heavy WIMPs, the available energy via the high velocity tail of the Maxwell-Boltzmann (M-B) distribution maybe adequate 
 \cite{EFL88} to allow  scattering  to low lying excited states of of certain targets, e.g. the 
$57.7~$keV for the $7/2^{+}$ excited state of $^{127}$I, the 39.6 keV for the first excited $3/2^{+}$ of $^{129}$Xe, the 35.48 keV for the first excited $3/2^+$ state of $^{125}$Te  and the 9.4 keV for the first excited $7/2^{+}$ state of $^{83}$Kr .

In fact we expect that the branching ratio to the excited state will be enhanced  in the presence of an  energy threshold for the detector , since only the total rate to the ground state transition  will be affected and reduced by the threshold, while this will have a negligible effect on the rate  to the WIMP-nucleus inelastic scattering. In  the present paper we  focus on studying the WIMP-nucleus inelastic scattering to the first excited $3/2^+$  state of $^{125}$Te, employing appropriate spin structure functions obtained in the context of  realistic shell model calculations . All isospin channels arising in various popular particle models are considered, but  it turns out that by  combining the proper particle and nuclear inputs, the isovector nucleon cross section is the most important. 
\section{The spin dependent WIMP-Nucleus scattering}
The spin dependent WIMP-Nucleus cross section is typically expressed in terms of the WIMP-nucleon cross section, which contains the elementary particle parameters entering the problem at the quark level. From the particle physics point of view the interaction of WIMPs with ordinary matter is given at the quark level by two amplitudes, one isoscalar $\alpha_0(q)$  and one isovector  $\alpha_1(q)$. In going to the nucleon level one must   transform these  two amplitudes by suitable
renormalization factors: 
\begin{itemize}
\item[i)] In terms of the
quantities $\Delta q$ prescribed by Ellis \cite{JELLIS}, namely\\$\Delta
u=0.78 \pm 0.02$, $\Delta d=-0.48\pm 0.02 $ and $\Delta s=-0.15
\pm 0.02$, i.e.
\begin{align} &a_0=a(q)\hspace{2pt}(\Delta u+\Delta
d+\Delta s)=0.15\hspace{2pt}a_0(q), \nonumber \\
& a_1=a_1(q)\hspace{2pt}(\Delta u-\Delta
d)=1.26\hspace{2pt} a_1(q),
\label{Eq:a0a1A}
\end{align}
where $a_i(q)$ and $a_i,\,i=0,1$ are the amplitudes at the quark and nucleon levels. In most applications  $\alpha_i(q)=1$.
\item [ii)] In terms  of a recent analysis \cite{QCDSF12,LiThomas15}, also found appropriate for pseudoscalar couplings to quarks \cite{ChengChiang12}: 
$\Delta u=0.84 $, $\Delta d=-0.43 $ and $\Delta s=-0.02$. Thus in general
\begin{align} &a_0=a_0(q)\hspace{2pt}(\Delta u+\Delta
d+\Delta s)=0.39\hspace{2pt}a_0(q), \nonumber \\
& a_1=a_1(q)\hspace{2pt}(\Delta u-\Delta
d)=1.27\hspace{2pt} a_1(q).
\label{Eq:a0a1B}
\end{align}
In the case of the $Z$-exchange mechanism, however, the the isoscalar contribution at the nucleon level is very small \cite{Thomas02}.
\end{itemize} 
It seems, therefore, that  the isovector component
behaves  as it  is expected for the axial current of weak interactions, while the
isoscalar component is suppressed, consistent with  European Muon Collaboration (EMC) effect \cite{EMC83,EMC95},  i.e. the fact that only a tiny fraction of the spin of the nucleon is coming from the spin of the quarks (see, e.g., an update in   a recent review \cite{SpinStr15}).
 In the latter case the isoscalar  is quite a bit larger. This may become of interest in the case of the inelastic transition to the first excited of  $^{125}$Te where the isoscalar contribution is favored by nuclear structure.

\begin{figure}
\begin{center}
 \begin{subfigure}[]
{
\includegraphics[width=0.45\textwidth]{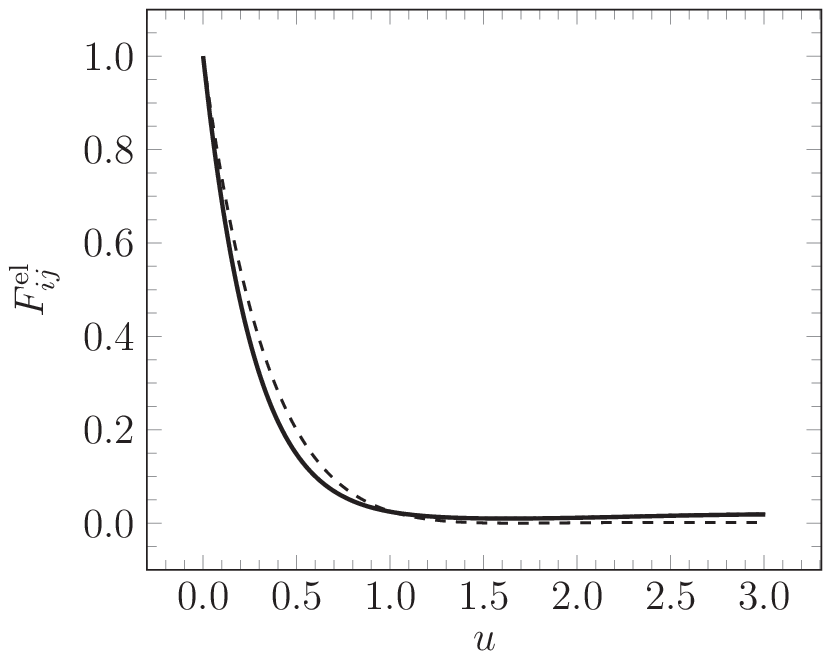}
}
 \end{subfigure}
 \begin{subfigure}[]
{
\includegraphics[width=.45\textwidth]{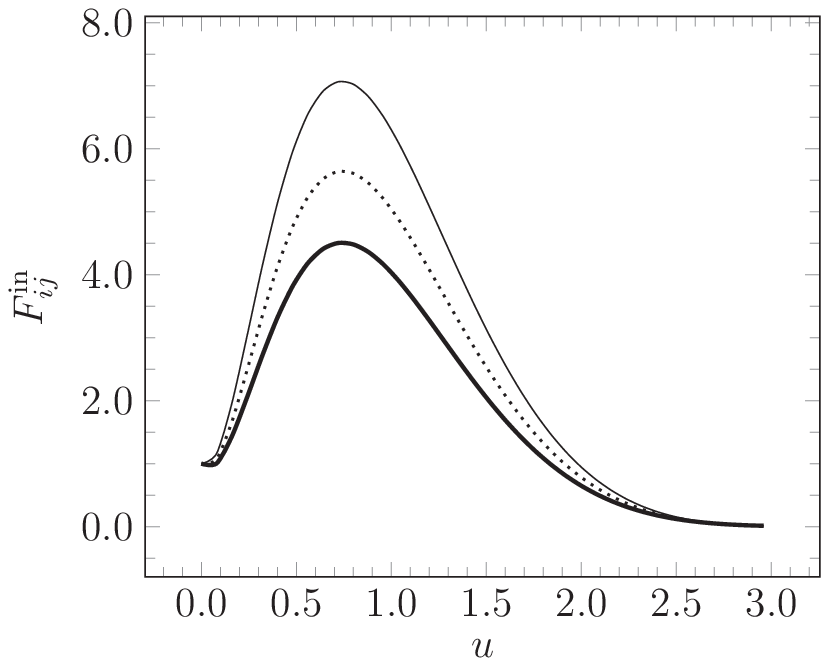}
}\end{subfigure}
\\
\caption{The elastic spin  structure functions $F^{e\ell}_{ij} ( u)$ for the isospin modes $i,j={11},{01},{00}$, as functions of $u=E_R/ Q_0$  in the  case of the target $^{125}$Te (a). For comparison we present the square of the form factor entering the coherent mode (dashed line). Note that the elastic structure functions, when normalized to unity at zero momentum transfer, are essentially independent of isospin.  We also show the inelastic spin structure functions $F^{in}_{ij}(u)$ (b). In this case the structure functions show substantial differences . In all cases the solid, dotted and thick solid curves  correspond to $F_{00},\, F_{01}$ and $F_{11}$ respectively}.
 \label{fig:FFspinTe}
\end{center}
\end{figure}

\begin{figure}
\begin{center}
\includegraphics[width=0.8\textwidth]{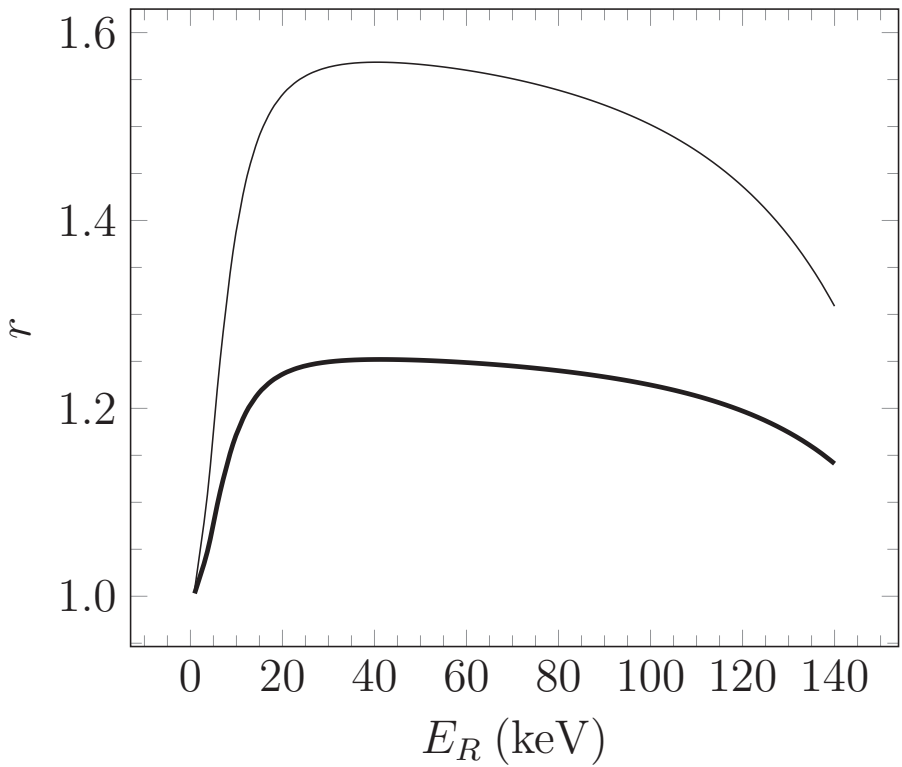}\\
\caption{The functions $r_{00}(E_R/Q_0)$ (thick solid line) and $r_{01}(E_R/Q_0)$ (solid line) entering the inelastic scattering to the first excited state of $^{125}$Te as a function of the recoil energy in keV. We see that, in the energy transfer region allowed for the  inelastic scattering of a 50 GeV WIMP mass,  the values of
$r_{00}$ and $r_{01}$ are about 1.5 and 1.2, i.e. the $F_{00}$ and $F_{01}$ are   larger than $F_{11}$.
 \label{fig:rFFspinTe}}
\end{center}
\end{figure}

From these amplitudes one  is able to calculate the isoscalar and isovector nucleon cross section. It is for this
reason that we started our discussion in the isospin basis and not the proton neutron representation preferred by some other authors.
 After that, within the context of a nuclear model, one can obtain the nuclear matrix element:
\beq
  |ME|^2 =a_1^2 S_{11}(u)+a_1 a_0 S_{01}(u)
  +a^2_{0}S_{00}(u),
\eeq
where     $S_{ij}$ are the spin structure functions, which depend on  the nuclear wave functions involved and the  energy transferred to the nucleus, indicated by  $u$  in dimensionless units which will   be appropriately defined below.  It  is useful to define the structure functions 
$$F_{ij}=\frac{S_{ij}}{\Omega_i \Omega_j}$$
with $\Omega_1,\,\Omega_0$ the isovector and isoscalar static spin matrix elements. Then the functions $F_{ij}$ take the value unity at zero momentum transfer. Furthermore these three  structure functions are approximately the same, in the case of the nuclei previously studied   \cite{DIVA00,DivVer13,VAPSKS15} as  well as in the case of elastic scattering  involving $^{125}$ Te, as it will be shown below . The event rate is given by Eq. (\ref{drdus}) of the Appendix A.
 \begin{align}
 \sigma_A^{\mbox{\tiny{spin}}}=\eta(u) 
  \frac{ \Omega^2_1\sigma_1(N)}{3},\,\eta(u)=\left ( 1+2\mbox{sign}(a_1 a_0)r_{01}(u)
   \frac{\Omega_0 }{\Omega_1}\sqrt{\frac{\sigma_0(N)}{\sigma_1(N)}}+ r_{00}(u)\frac{\Omega^2_0 \sigma_0(N)}{ \Omega^2_1\sigma_1(N)}\right),
\label{Eq:nosf}
\end{align}
 where $\sigma_0(N)$ and $ \sigma_1(N)$ are the elementary nucleon isoscalar and isovector  cross sections, the factor $F_{11}(u)$ is included separately and
	$$r_{01}(u)=\frac{F_{01}(u)}{F_{11}(u)},\,r_{00}(u)=\frac{F_{00}(u)}{F_{11}(u)}$$
	Then  if the  the isoscalar nucleon cross section is  suppressed, the above expression simplifies to: 
	 \beq
  \sigma_A^{\mbox{\tiny{spin}}}=
  \frac{ \Omega^2_1\sigma_1(N)}{3},
  \label{Eq:nosf0}
  \eeq
	In most calculations performed so far the quantity $\eta(u)$ in Eq. \ref{Eq:nosf} has been found to be essentially independent of $u$. The obtained results for the structure functions relevant, e.g., for $^{125}$Te are exhibited in Fig. \ref{fig:FFspinTe}.
We see that in the case of elastic scattering (a) the three spin functions $F_{ij}(u)$ are almost identical. The square of the form factor relevant for coherent event rates (dashed) curve differs only slightly from the spin structure functions. The structure functions relevant for the inelastic scattering  show significant differences (the solid, dotted and thick solid curves correspond to $F_{00},\, F_{01}$ and $F_{11}$ respectively). The fact that $\eta(u)$ is no longer a constant is better illustrated by exhibiting the ratios $r_{00}$ and $r_{01}$ in Fig. \ref{fig:rFFspinTe}.



The above expressions look complicated. In most existing models, however, the situation is as follows:
\begin{itemize}
\item The process mediated by  $Z$ exchange, Supersymmetric models and  Kaluza Klein theories in models with Universal Extra Dimensions \cite{OikVerMou,ST02a,ST02b} involving  heavy neutrinos as WIMPs. Then Eqs (\ref{Eq:a0a1A}) and (\ref{Eq:a0a1B}) yield
$$\frac{\sigma_0}{\sigma_1}=\left \{ \begin{array}{cc}\left(\frac{0.15}{1.26}\right)^2=0.014&\mbox{ case i) }\\\left(\frac{0.39}{1.27}\right)^2=0.094&\mbox{ case ii) }\\ \end{array}\right .$$
The absolute scale of the nucleon cross section depends on the details of the model.
\item Kaluza-Klein theories in models with Universal Extra Dimensions \cite{OikVerMou,ST02a,ST02b} the WIMP happens to be a vector boson. In this case one finds:
$$a_0(u)=\frac{17}{18},\,a_0(d)=a_0(s)=\frac{5}{18}$$
Thus for case i) we get:
$$ a_0=(17/18)\times0.78-(5/18)\times0.48-(5/18)\times 0.15=0.56,\,a_1=(17/18)\times0.78+(5/18)\times 0.48=0.87,$$
while for case ii)
$$ a_0=(17/18)\times0.84-(5/18)\times0.43-(5/18)\times 0.02=0.69,\,a_1=(17/18)\times0.84+(5/18)\times 0.43=0.91,$$
Thus
$$\frac{\sigma_0}{\sigma_1}=\left \{ \begin{array}{cc}\left(\frac{0.56}{0.87}\right)^2=0.41&\mbox{ case i) }\\\left(\frac{0.69}{0.91}\right)^2=0.53&\mbox{ case ii) }\\ \end{array}\right .$$
i.e. the isoscalar contribution is sizable.
\item  The WIMP is a spin 3/2 particle \cite{SavVer13}. In this case only the isovector contribution exists, leading to $\sigma_1(N)\approx 1.7 \times 10^{-38}$cm$^{-2}=1.7 \times 10^{-2}$pb.
\end{itemize}
In all the above cases the ratio  of the elementary amplitudes is positive.

\begin{figure}
\begin{center}
 \begin{subfigure}[]
{
\includegraphics[width=0.4\textwidth]{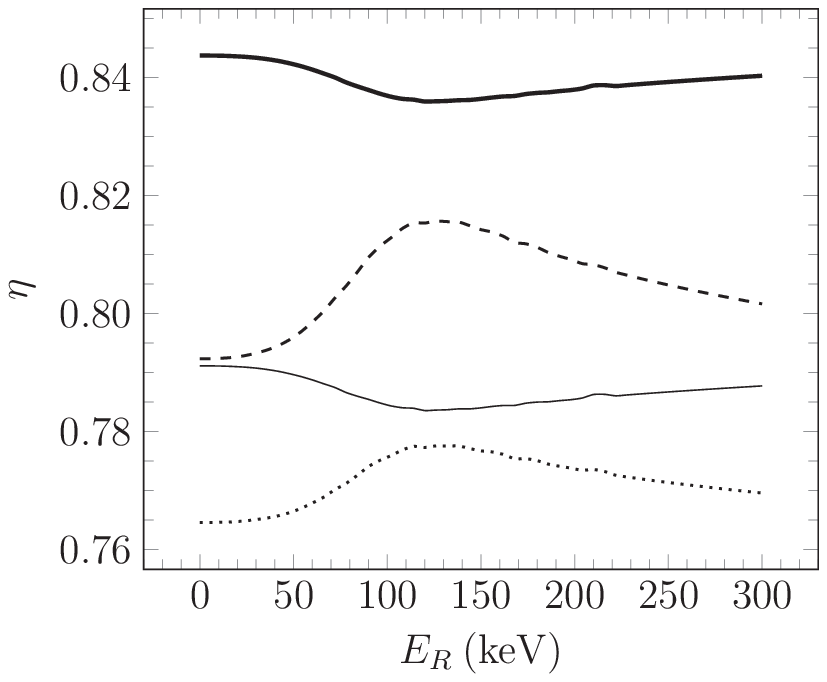}
}
\end{subfigure}
\begin{subfigure}[]
{
\includegraphics[width=0.4\textwidth]{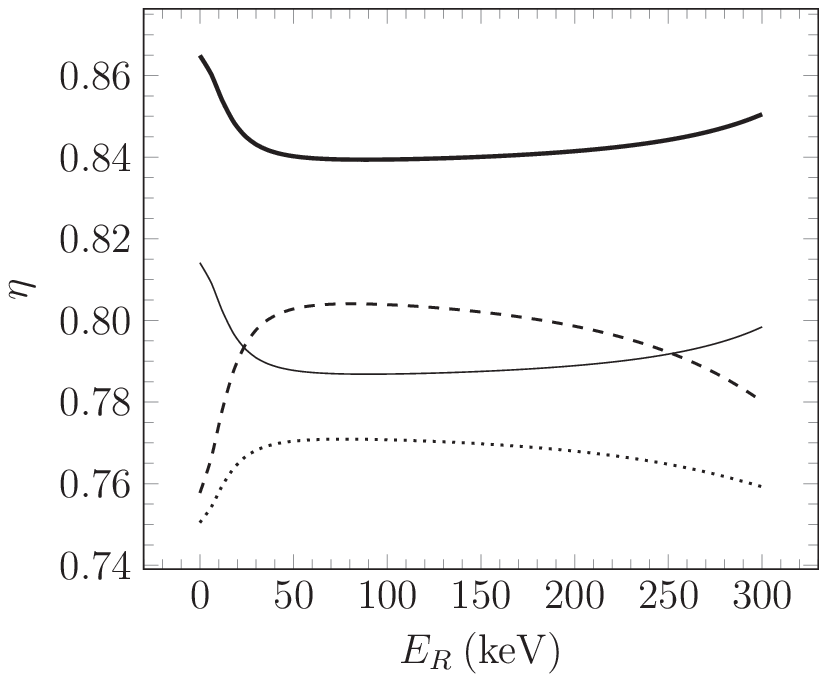}
}
\end{subfigure}
\\
\caption{The functions $\eta(E_R/Q_0)$  for a 50 GeV WIMP  in the  case of the target $^{125}$Te as a function of the recoil energy in keV. For elastic scattering (a) and the inelastic scattering (b).  The thick solid, the solid, the dotted and the dashed curves correspond to $\sigma_0/\sigma_1=0.014,\,0.094,\,0.41 $ and 0.53 respectively. The presence of the isoscalar tends to decrease the rates. This is understood, since the relevant nuclear matrix elements in both cases have opposite sign, but the actual reduction is mild, no more than 10$\%$  reduction in the total rate.
 \label{fig:EtaFFspinTe}}
\end{center}
\end{figure}

As an example we plot in Fig. \ref{fig:EtaFFspinTe} the function $\eta(u)$ corresponding to a WIMP mass of $50$ GeV, for  the elastic as well as the inelastic scattering to the first excited state of $^{125}$Te. We see that
the tendency of the isoscalar component is to reduce the rates.
\begin{table}[htbp]
\begin{center}
\caption{ The parameter $\eta$ entering the elastic scattering for various nuclei. This parameter renormalizes the dominant isovector nucleon cross section due to the presence of isoscalar contributions}
\label{tab:eta}
\begin{tabular}{ ||c|  c  c  c c| |} 
\hline
\hline
&&&$\sigma_0/\sigma_1$&\\
&0.04& 0.094& 0.41& 0.53\\
\hline
A=127&1.284& 1.479& 2.284& 2.544\\
A=129&0.842& 0.789& 0.767& 0.798\\
A=83&0.838& 0.785& 0.773& 0.808\\
A=125&0.844& 0.791& 0.765& 0.792\\
\hline
\hline
\end{tabular}
\end{center}
\end{table}
In the inelastic case, since the low energy transfer region does not contribute, we get a contribution essentially independent of the energy transfer  leading to a decrease in the rates  by  an overall  constant factor between  5$\%$ and 10$\%$. In the elastic case the functions $\eta(u)$ depend  mildly on the energy, but the low energy transfer  is now favored  and we expect a decrease  from around 4$\%$ to  9$\%$. The decrease in the case of this target  is  understood, since the relevant nuclear matrix elements in both cases have opposite sign.
	
	Since the experiments are analyzed in the proton-neutron representation we transform the above results to this basis. Thus $\sigma_p/\sigma_1=(1/2)(1+\sigma_0/\sigma_1)$,  $\sigma_n/\sigma_1=(1/2)(-1+\sigma_0/\sigma_1)$. We thus get 
	$$ \sigma_p=(0.520, 0.547, 0.705, 0.765)\sigma_1,\,\sigma_n=(-0.480, -0.453, -0.295, -0.235)\sigma_1$$
	in the order given in table \ref{tab:eta}. The scale $\sigma_1$ depends on the particle model.
	We should mention at this point that there exist some experimental limits, namely for $^{129}$Xe and $^{131}$Xe \cite{Xenon100.11} and $^{19}$F \cite{COUPP12,SIMPLE12,PICASSO12,PICASSO09,PICASSO08}. From the Xe  data   a limit is extracted on the elementary neutron SD cross section of $\sigma_n=2\times 10^{-40}$cm$^2=2\times 10^{-4}$pb and $\sigma_p=2\times 10^{-38}$cm$^2=1.0\times 10^{-2}$pb for the proton SD cross section, while  from the $^{19}$F target  a slightly smaller limit is extracted  on the proton SD cross section,  $\sigma_p=1\times 10^{-38}$cm$^2=1.0\times 10^{-2}$pb. These limits were based on nuclear physics considerations, namely the nuclear spin matrix elements in the proton-neutron representation. This explains the difference of the two limits extracted from the Xe data.
	On the other hand, if the elementary amplitude is purely isovector, these limits would imply  $\sigma_1=\sigma_p-\sigma_n=1.02\times 10^{-2}$pb.  We should mention in passing that the nuclear matrix elements for $^{19}$F are expected to be much more
	reliable \cite{DIVA00,DivVer13}.  Actually the problem of extracting the  nucleon cross sections from the data will remain open until all the three nucleon cross sections (scalar, proton spin and neutron spin) can be determined along the lines previously suggested \cite{CannVerGom11}, after sufficient experimental information on at least three suitable  targets becomes available.
	In the present work we will not commit ourselves to any particular model, but for orientation purposes we will use the value  mentioned above, i.e. \cite{SavVer13}
$\sigma_1(N)\approx 1.7 \times 10^{-38}$cm$^{-2}=1.7 \times 10^{-2}$pb, even though it tends to be on the larger side, which of course gets renormalized by the factors $\eta$ given in table \ref{tab:eta}.
\section{Shell-Model Interpretation of the spin structure functions - Aplication in the case of the  $^{125}$T\lowercase{e}  nucleus}
\label{discuss_SM}
 A complete calculation of the relevant spin structure should be performed
along the lines  previously done for other targets \cite{Ress,PITTEL94,VogEng,IACHELLO91,NIKOLAEV93,SUHONEN03,Nsuhonen} 
and more recently \cite{MeGazSCH11,MeGazSCH12,BKKLMRA13} in   the shell model framework. 

 We will not concern ourselves with other simplified  models, e.g schemes of deformed rotational nuclei
 \cite{EngVog00}.

 A  summary of some nuclear Matrix Elements (MEs) involved in elastic and inelastic scattering can be found in Refs \cite{SavVer13}, \cite{VerEjSav13}.
Here we concentrate on the shell-model calculations  carried out for $^{125}$Te in the
50-82 valence shell composed of the orbits $1g_{7/2}$, $2d_{5/2}$,
$1h_{11/2}$, $3s_{1/2}$, and $2d_{3/2}$  with
SN100PN interaction due to Brown {\it et al.}~\cite{SMCal96,SMCal05}. This interaction has four parts: 
neutron-neutron, neutron-proton, proton-proton and Coulomb repulsion between the protons.
The single-particle energies for the neutrons
are -10.6089, -10.289, -8.717, -8.694, and -8.816 MeV
for the $1g_{7/2}$, $2d_{5/2}$, $2d_{3/2}$, $3s_{1/2}$, and $1h_{11/2}$ orbitals, respectively,
and those for the protons are 0.807, 1.562, 3.316, 3.224,
and 3.605 MeV. In the present calculation we slightly modified the single-particle energy of the $\nu2d_{3/2}$ orbital 
from -8.717 MeV to -8.017 MeV (changing by 700 keV).
We performed the shell-model calculation using truncation because of the large matrix dimension involved in the
present calculation; therefore, we put two valence protons in the $\pi1g_{7/2}$ and $\pi2d_{5/2}$ orbits,
for neutrons we completely filled the $\nu1g_{7/2}$ and $\nu2d_{5/2}$ orbits, and put at least six neutrons
in the $\nu1h_{11/2}$ orbit.  

Results for a few low-lying states are compared in  Table \ref{tab:en}.
We performed the calculation with the shell model code NuShellX \cite{NuShellX}. 
The present calculation predicts the negative-parity spectrum at very low energy and the ground state becomes $11/2^-$. The order and relative energies of the important positive-parity states were quite nicely reproduced, however. The present calculation predicts  $3/2^+$ at 49 keV with respect to $1/2^+$, while the corresponding experimental
value is 35.5 keV. The configurations of the $1/2^+$ and $3/2^+$ states are $\nu 3s_{1/2}$ and $\nu2d_{3/2}$, respectively, while the configurations of $5/2^+$ and  $7/2^+$ are mixed character of both $\nu 3s_{1/2}$ and $\nu2d_{3/2}$ orbits.
We have also calculated electromagnetic properties which are shown in Tables  \ref{tab:be2} and \ref{tab:qm}.
The  calculated $B(M1)$($3/2^+$ $\rightarrow$ $1/2^+$) value is 0.00562 W.u. (with $g_s^{eff}$ = $g_s^{free}$),
while the corresponding experimental value is $B(M1)$=0.0226(4) W.u.

\begin{table}[htbp]
\begin{center}
\caption{ Calculated low-lying positive parity states of $^{125}$Te.}
\label{tab:en}
\vspace{2mm}
\label{tab:table3a}
\begin{tabular}{ c  c  c  c  } \hline
\hline
 $E_x (\mathrm{exp})$   & $E_x (\mathrm{SM})$   & $J^\pi$ &  Configuration   \\ \hline
              0 &           0   & $1/2^+$ &      $\nu s_{1/2}$ (68\%)               \\ 
              35.5 &          49   & $3/2^+$ &      $\nu d_{3/2}$ (53\%)                 \\ 
              402 &           459   & $7/2^+$ &  mixed [ $2^+$$\otimes$ $\nu d_{3/2}$ (57\%) + \\
                  &                 &         &  $2^+$$\otimes$ $\nu s_{1/2}$ (30\%)  ]              \\ 
             463 &           449    & $5/2^+$ &  mixed [$2^+$ $\otimes$ $\nu d_{3/2}$ (39\%) + \\
                  &                 &         &    $2^+$ $\otimes$ $\nu s_{1/2}$ (16\%) ]                 \\ 

\hline
\hline
\end{tabular}
\end{center}
\end{table}

\begin{table}[htbp]
\begin{center}
\caption{ $B(E2)$ and $B(M1)$ in W.u. Effective charges 
  $e_p=1.5$ $e_n=1.0$ were used for the $B(E2)$ calculation, while for the 
  $B(M1)$ calculation the bare g-values $g_s^{eff}$ = $g_s^{free}$ were used. Experimental values were taken from
  Ref.~\cite{nndc}.}
\label{tab:be2}
\vspace{2mm}
\label{tab:table3b}
\begin{tabular}{ c  c  c  c  } \hline
\hline
   & Transitions  &Expt. &SM  \\ \hline
$^{125}$Te & $B(E2;3/2^+ \rightarrow 1/2^+)$ & 11.9(24) & 6.03 \\ 
$^{125}$Te & $B(M1;3/2^+ \rightarrow 1/2^+)$ & 0.0226(4) & 0.00564 \\ 
\hline
\hline
\end{tabular}
\end{center}
\end{table}

\begin{table}[htbp]
\begin{center}
\caption{ Comparison of experimental and calculated electric quadrupole moments (the effective
charges $e_p$=1.5, $e_n$=0.5 are used in the calculation) and magnetic moments (with $g_s^{eff}$ = $g_s^{free}$).}
\label{tab:qm}
\vspace{2mm}
\label{tab:table3}
\begin{tabular}{ c  c  c  c  c c } \hline
\hline
& & Q (eb) & &$\mu$ ($\mu_N$) & \\ 
\cline{3-6}
   &    &Expt. &SM &Expt. & SM \\ \hline
$^{125}$Te& $1/2^+$ &  &  &  -0.8885051(4) & -1.598  \\ 
     & $3/2^+$ & -0.31(2) & -0.18 &  +0.605(4)  & +0.950 \\ \hline
\hline
\end{tabular}
\end{center}
\end{table}


 The static spin matrix elements (MEs) obtained from this calculation  are
$$ \Omega_0=1.456,\,\Omega_1= -1.502  \mbox{ for elastic transitions},$$
$$ \Omega_0= -0.157,\,\Omega_1= 0.196 \mbox{ for inelastic transitions}.$$
Looking at these matrix elements we notice that the nuclear matrix elements relevant for the inelastic scattering are, as expected, smaller than those entering the elastic scattering, but they are not suppressed as much as those relevant for the target $^{83}$Kr \cite{VAPSKS15},  the latter  favored kinematically due to its lower excitation energy. We also notice that
the nuclear structure does not favor the isoscalar contribution  to overcome the suppression  of the corresponding  isoscalar amplitude due to the structure of the nucleon.

 \section{transitions to excited states}
The expression for the elastic differential event rate is well known, see e.g. \cite{SavVer13},\cite{DivVer13}. For the reader's convenience, and to establish our notation, we briefly present the essential ingredients in the Appendix A.  We should mention  that, as can be seen in the appendix,  the difference in the shape of the spectrum between the coherent and the spin induced cross section comes from nuclear physics. Since the square of the form factor and the elastic spin structure functions have approximately similar shapes, we expect the corresponding differential shapes to be the same, and only the scale to be different.

Transitions to excited states are normally energetically suppressed, except in some odd-$\emph{A}$ nuclei that have low lying excited states . Then spin mediated transitions to excited states are possible. The most favorable are expected to be those that, based on the total angular momentum and parity of the states involved, appear to be  Gamow-Teller like transitions (see Table \ref{tab:ej1}).
\subsection{Isotope considerations}
Transitions to the first excited states via the spin dependent (SD) WIMP-nucleus interaction can occur in the case of some odd-$\emph{A}$ targets, if the relevant excitation energy is $E \le 100 $ keV. This is due to the high velocity tail of the Maxwell-Boltzman (M-B) distribution, so that a reasonable amount of the WIMP energy may be transferred to the recoiling nucleus.

Possible odd mass nuclei involved in DM detectors used for WIMP searches are state in $^{127}$I at 57.6 keV,  the 39.6 keV state in $^{129}$Xe and the 9.5 keV $^{83}$Kr. The spin excitations of these states are not favored, because the dominant components of the relevant wave functions  are characterized by $\Delta \ell\ne 0$, i.e. $\ell$ forbidden transitions.  Nevertheless the spin transitions are possible, due to the small components  as seen from the M1 $\gamma $ transition rates. Other possibilities are, of course, of interest, e.g. 
$^{125}$Te is a possible experimental candidate, as will be discussed below. 
 Thus it is quite realistic to study the inelastic excitations in this nucleus in the search for  WIMPs via the SD interaction. 

In fact the experimental observation of the inelastic excitation has several advantages. The experimental feasibility in the case of the $^{125}$Te target is discussed in section \ref{sec:ExperimentalFeasibility}.

\begin{table} [h]
\caption{Inelastic spin excitations of experimentally interesting targets
 $A$ : natural abundance ratio, $E$ : excitation energy, $J_i$: ground state spin parity, $J_f$ : excited state spin parity, and  $T_{1/2 } $: half life. 
}
\label{tab:ej1}
\begin{center}
\begin{tabular}{cccccc} 
  \\ 
\hline
Isotope &   $A(\%)$ &   $E$ (keV) &   $J_i$ &   $J_f$ &  $T_{1/2} $(ns) \\        
\hline
$^{83}$Kr    &       11.5  &        9.5  &    9/2$^+$  &  7/2$^+$ &    154\\  
$^{127}$I    &       100  &        57.6  &    5/2$^+$  &  7/2$^+$ &    1.9 \\  
$^{129}$Xe &       26.4  &      39.6  &    1/2$^+$  &   3/2$^+$ &     0.97\\ 
$^{125}$Te &       7.07 &      35.5  &    1/2$^+$  &   3/2$^+$ & 1.48  \\     

 \hline
\end{tabular}
\end{center}
\medskip 
\end{table}  

\subsection{Kinematics}

The evaluation of the differential rate for the inelastic transition proceeds in a fashion similar to that of the elastic case discussed above, except:
\begin{itemize}
\item The transition spin matrix element must be used.
\item The transition spin response function must be used.
For Gamow-Teller like transitions, it does not vanish for zero energy transfer. So it can be normalized to unity, if the static spin value is taken out of the MEs.
\item The kinematics is substantially modified compared to that of the elastic case. This, however, has been previously discussed\cite{VAPSKS15} and we will not  further elaborate.
\end{itemize}
We only mention that  differential event rate for inelastic scattering takes a form similar to that of the elastic scattering except that 
 $$\Omega_1\rightarrow \Omega_1^{\mbox{\tiny{inelastic}}},\, F_{11}(u)\rightarrow F_{11}(u)^{\mbox{\tiny{inelastic}}},\quad\Psi_0(a\sqrt{u})\rightarrow \Psi_0\left(a\frac{ u+u_0}{\sqrt{u}}\right ).$$
where the function $\Psi_0 $  as well as $a$ and  $u_0$ and is defined in the appendices A and  B.
\section{Some results}
For purposes of illustration we employ the  nucleon cross section of $1.7\times 10^{-2}$pb  obtained in a recent 
work \cite{SavVer13}, without committing ourselves to this or any other  particular model. Another input is the WIMP density in our vicinity, which will be taken to be 0.3 GeV cm$^{-3}$. Finally the velocity distribution with respect to the galactic center will be assumed to be a  M-B with a characteristic velocity $\upsilon_0=220$km/s and an upper cut off (escape velocity) of $2.84 \upsilon_0$.
 \subsection{Results for the differential event rates} 
\begin{figure}
\begin{center}
 \begin{subfigure}[]
{
\includegraphics[width=0.7\textwidth]{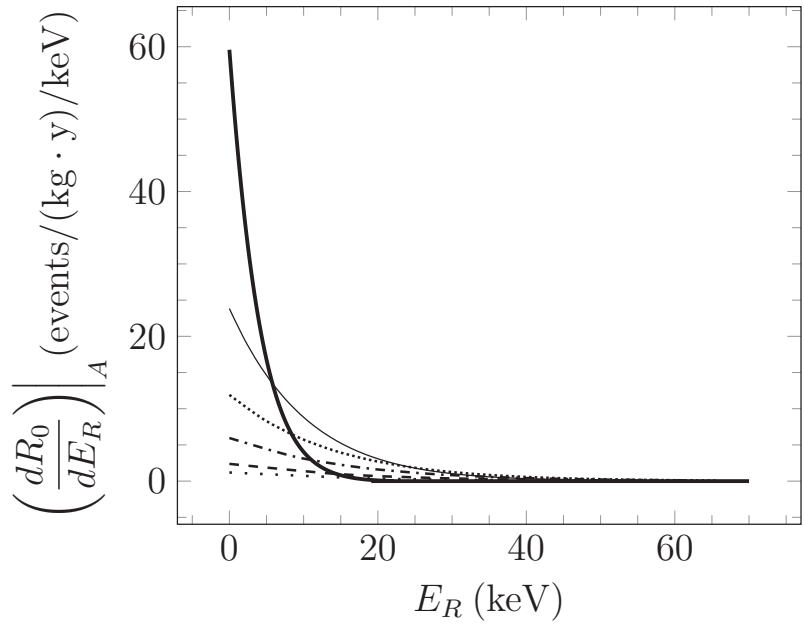}
}\\
\end{subfigure}
\begin{subfigure}[]
{
\includegraphics[width=0.7\textwidth]{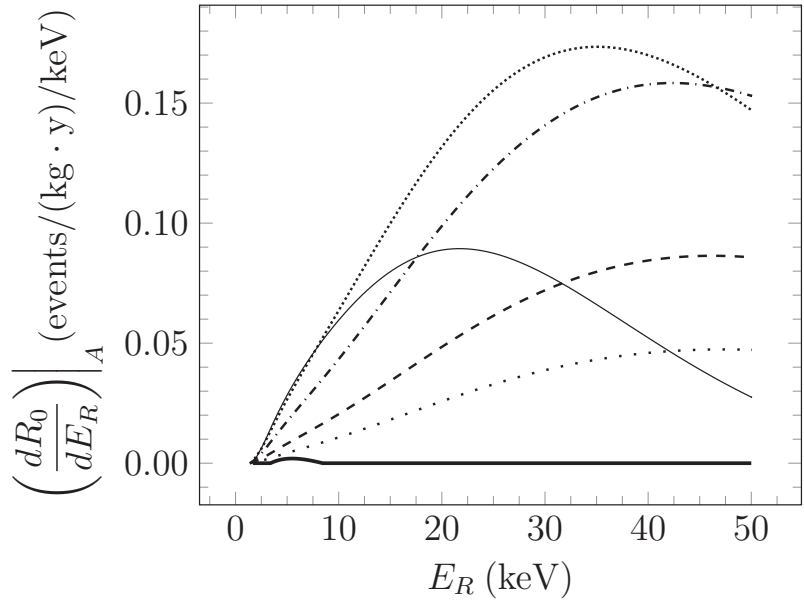}
}
\end{subfigure}
\\
\caption{ The energy spectrum  for WIMP-$^{125}$Te elastic scattering (a) and that for the inelastic scattering to the first excited state at 35.5 keV  (b). The thick solid, solid, dotted, dash-dotted, dashed and large dotted curves correspond to  WIMP masses of 20, 50, 100, 200, 500 and 1000 GeV respectively. In this case the quenching factor is unity.}
\label{fig:dRdQTe}
\end{center}
\end{figure}

The differential event rates, perhaps the most  interesting from an experimental point of view,  depend on the WIMP mass, but we  only present them for some select WIMP masses.
Our results for the  elastic differential rates for typical WIMP masses are  exhibited in Fig. \ref{fig:dRdQTe}a. For comparison we present the differential event rates for transition to the excited state in Fig. \ref{fig:dRdQTe}b.
The large energy signal is obtained by summing the nuclear recoil signal and the $\gamma $ ray signal. It is given as  
\begin{equation}
E({\rm ex})=E_{\gamma} + Q(E_R({\rm ex}))E_R({\rm ex}), 
\end{equation}
where $E_R({\rm ex})$ is the nuclear recoil energy,  $E_{\gamma }$ is the excitation
energy and $Q(E_R({\rm ex}))$ is the quenching factor for the recoil energy signal. It must be determined for each target and detector experimentally.  In the present study, however, we assume that the detector is going to be a bolometer, which is characterized by a quenching factor approximately unity. So in this case the inelastic channel cannot benefit as much from the fact that the quenching factor tends to suppress the ground state transition in the presence of significant  energy threshold as found previously \cite{VerEjSav13,VAPSKS15}.
\subsection{Total rates}
\begin{figure}
\begin{center}
\begin{subfigure}[]
{
\includegraphics[width=0.6\textwidth]{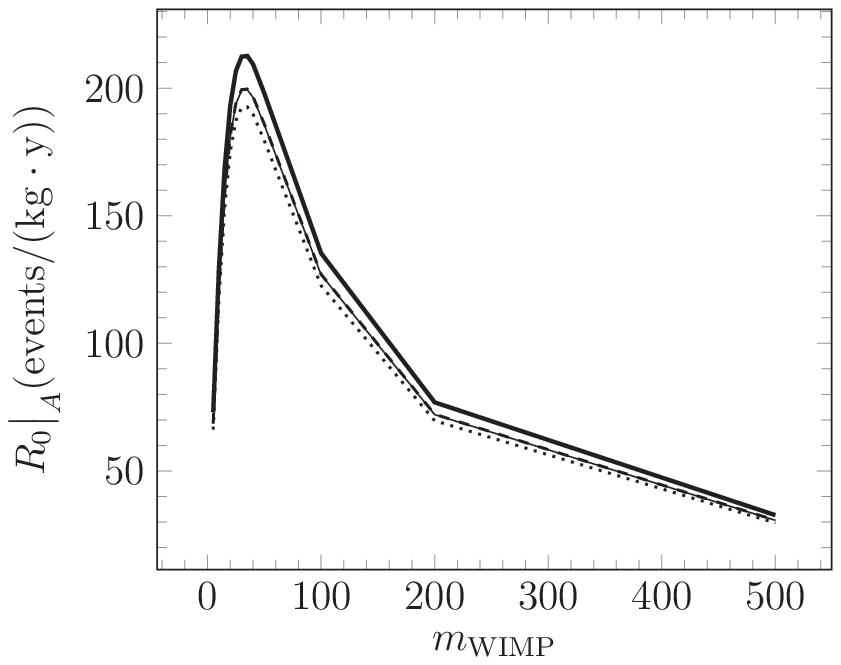}
}
\end{subfigure}[]
\\
\begin{subfigure}[]
{

\includegraphics[width=0.6\textwidth]{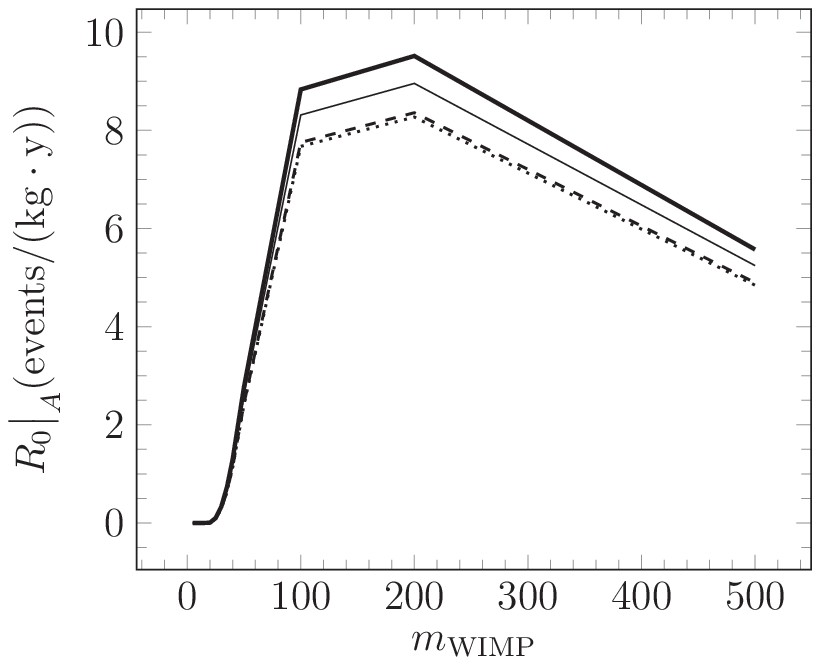}
}
\end{subfigure}
\\
\caption{ The calculated	time	average	total	rates	in	events/kg/y,	as	a	function	of	
WIMP	mass	in	GeV,	for	elastic scattering	(a)	and	inelastic	scattering (b),	assuming	a	
zero-energy	threshold.	The	thick	solid,	solid	dotted	and	dash	curves	correspond	
to	 $\sigma_0 / \sigma_1$ =0.04,	0.094,	0.41	and	0.53,	respectively.	 The value of $\sigma_1 \times 10^{-3}$pb was employed. Note that the location of the maximum  rate  in the case of the inelastic scattering has shifted compared to that of the  ground state. 
 \label{fig:totR0Te}}
\end{center}
\end{figure}

From expressions (\ref{Eq:Trates}) and (\ref{Eq:Tratec}) of the Appendix A, we can obtain the total rates.
The total rates obtained assuming zero energy threshold are exhibited in Fig. \ref{fig:totR0Te} as functions of the WIMP mass. We also exhibit the dependence of the total rates on the isoscalar nucleon elementary cross section. With the employed  isovector  nucleon cross section of $1.7\times 10^{-2}$pb the total rates expected are quite large. Due to favorable nuclear physics the total rate for inelastic scattering is quite high, about 10$\%$ of that to the ground state regardless of the assumed nucleon cross section.

\section{ Experimental aspects of inelastic nuclear scattering rates} 
In this section, we discuss experimental aspects of the spin dependent (SD) WIMP-nucleus  interaction involving inelastic nuclear scattering. So far SD and SI WIMP interactions with nuclei have been studied experimentally by measuring nuclear recoils in elastic scattering. 

SD interactions may show fairly appreciable cross sections for inelastic spin excitations, as shown in previous sections. Experimentally, inelastic nuclear excitations provide unique opportunities
for studying WIMPs exhibiting SD interactions with hadrons.  
 Experimentally, inelastic nuclear excitations provide unique opportunities
for studying SD  rates for WIMP-nuclear interactions.  Inelastic excitations can, in principle, be studied by two ways:  A singles measurement of both the nuclear recoil energy $E_R$ and the decaying $\gamma $-ray energy $E_{\gamma }$ in one detector, and B a  coincidence measurement of the nuclear recoil and the $\gamma $-ray in two separate detectors in a fashion discussed in the earlier analysis \cite{VerEjSav13,VAPSKS15}. In the case of $^{125}$Te we will consider only option A (see \cite{VerEjSav13,VAPSKS15} for details).
 
 We should mention that a search for inelastic WIMP nucleus scattering on $^{129}$Xe using  data from the XMASS-I experiment has recently appeared \cite{Xmass14}. These authors set an upper limit of 3.2 pb  for the inelastic nucleon cross section. This large value has nothing to do with the particle model, but it is a manifestation of the unfavorable kinematics involved in the inelastic case.
\section{Experimental Feasibility}
      \label{sec:ExperimentalFeasibility}                    

            One of the main purposes of the research described in this article is to explore the feasibility of an experiment to search for Cold Dark Matter via  inelastic excitation of a nuclear target. The isotope $^{125}$Te was chosen  for two reasons: first, it has a fairly low energy M1 transition from the 3/2$^{+}$ first excited state at 35.5-keV to the 1/2$^{+}$ ground state; second, TeO$_2$ has been proven to be an excellent detector material for high-resolution bolometers as demonstrated experimentally in the CUORICINO \cite{Arnaboldi08,SuvTch93,Artusa15},
						 $0\nu\beta\beta$-decay experiments. One disadvantage of this choice is that the natural abundance of $^{125}$Te is only 7.07$\%$, which would require isotopic enrichment if it becomes necessary to enhance sensitivity. However, it has been demonstrated that Te can be enriched to more than 90$\%$ in $^{125}$Te by the gas centrifuge technique by converting the Te into tellurium hexafluoride (TeF$_6$), which is stable and is the gas used in gas centrifuges to enrich Te isotopes. A 10-kg sample, enriched to 93$\%$ in $^{130}$Te was produced for the University of South Carolina by the Electro-Chemical Plant (ECP) in Zelenogorsk, Russia  \cite{Arnaboldi08}, and successfully converted into TeO$_2$ bolometers by the Shanghai Institute of Ceramics of the Chinese Academy of Science (SICCAS) in Shanghai, China. The enrichment in $^{125}$Te can be done by the same procedure.
              The series of experiments performed by the CUORICINO and CUORE Collaborations has clearly demonstrated that arrays of many kilograms of TeO$_2$ can be operated as bolometers at ~8-10 mK for years. The CUORE detector nearing completion in the Laboratori Nazionali del Gran Sasso (LNGS), in Assergi, Italy will operate ~740-kg of TeO$_2$ bolometers for a planned five year period to search for the neutrino-less double-beta decay of $^{130}$Te. CUORE contains bolometers fabricated from natural abundance Te of which 33.08$\%$ $^{130}$Te. Therefore CUORE will contain ~200-kg of $^{130}$Te, and 43.75-kg of $^{125}$Te. An array similar to CUORE, but with bolometers enriched to 90$\%$ in $^{125}$Te, would contain ~562-kg of $^{125}$Te.
							
               The absolute value of the rates shown in Fig. \ref{fig:totR0Te} depends, of course, on the magnitude of the elementary isovector spin nucleon cross section, which is not known and  may have been somewhat overestimated.  The ratio, however,  of the inelastic to elastic rates is unaffected by this choice.  Anyway according to Fig. \ref{fig:totR0Te}, a 25-GeV WIMP will excite ~0.1 nuclei per kg-1y-1 or about 22 in a five-year data collection with CUORE. However, in a similar detector enriched to 95$\%$ in $^{125}$Te, 25-GeV WIMPS would excite ~290 nuclei.  One could easily imagine enriching only the 7-central CUORE towers, or 364 bolometers, which would result in ~ 540 predicted events. In addition, these seven inner towers will be shielded by the outer towers, and the background will be significantly reduced. According to the CUORE-0 article \cite{Alfonso15}, the reduction of background of the entire 19-tower array over that of CUORE-0 is predicted to be about a factor of 5. It should be noted that there are no data to support this conclusion; however, there were Monte-Carlo simulations that do support it \cite{Alfonso15}.
								
 While recently there is great interest in the lighter WIMPS, the picture is far more encouraging for the heavier WIMPS according to 	Fig. \ref{fig:totR0Te}. For 50-GeV WIMPS, the predicted rate is 2.77, 2.61, 2.40 and 2.43 per kg-y for $\sigma_0/\sigma_1$ values 0.04, 0.094, 0.41 and 0.53, respectively. In  the natural abundance CUORE  these numbers  become 108, 102, 94  95 and 34  events per year.  For 100-GeV WIMPS , our prediction being   8.88, 8.31, 7.67, and 7.75 per kg-y in the above order, these rates should be multiplied by 3.2 for all possible isoscalar factors ( it is the same since the rates in this region increase  almost linearly). In fact, a search of the CUORE-0 and CUORICINO data for 50 or 100-GeV WIMP signals could be interesting. The predicted number of events in each is ~6 and 20 events for 50 and 100-GeV WIMPS respectively, using the central value of 0.094. 
          In a previous article \cite{VAPSKS15}, a similar analysis was done for $^{83}$Kr. This isotope was chosen because large Noble liquid detectors are very successful in Cold Dark Matter searches, and the excitation energy to the first excited state is 9.4-keV. However, the difficulty in reducing the level of $^{85}$Kr sufficiently would present a real challenge. In addition, the excitation rates for $^{125}$Te are significantly higher, while the radioactive backgrounds in TeO$_2$ bolometers have been well studied by the CUORE Collaboration. On the other hand, the isotopic enrichment of Kr gas is far less costly than that of Te, because of the requirement to use TeF$_{6}$ to form a stable gas. Finally, the first search will be possible with CUORE, which is scheduled to take data in 2016 \cite{Artusa15}.

\section{Concluding remarks}

  WIMPs have extensively been studied, so far, by measuring elastic nuclear recoil involving both  SI and SD interactions.  The SI elastic scattering of WIMPs is coherent, thus the cross section is enhanced by the factor $A^2$, with $A$ being the nuclear mass number.  On the other hand, the elastic spin induced (SD) cross section of WIMPs  is typically assumed to be smaller by 2-3 orders of magnitude than that for SI WIMPs, if the relevant elementary nucleon cross sections are similar, because the spin induced rates do not depend on $A^2$, i.e. they do not exhibit coherence. It may, however, compete with the coherent scattering in models in which the spin induced nucleon cross section is much larger than the one due to a scalar interaction. We have seen that there exist viable particle models of this kind. In such cases the inelastic WIMP-nucleus scattering  becomes important.
 
 Indeed the inelastic scattering via SD interaction provides a new opportunity for detecting WIMPs via the SD interaction. Experimentally, the observation of both the nuclear recoil energy and the $\gamma $ ray following the  de-excitation of the populated state results in a large energy signal of the $E_{\gamma }$ and  a sharp rise of the energy spectrum at around $E_{\gamma }$.  This is true especially in the presence of quenching since  $E_{\gamma }$ is not quenched, whereas the elastic channel is quenched. How much smaller is the SD inelastic cross section depends on how close to  Gamow-Teller like is the inelastic transition.
In the case of $^{125}$Te the nuclear matrix elements are an order of magnitude larger than those of $^{83}$Kr, but it cannot benefit from the threshold energy suppression of the elastic transition, because in this case the quenching factor is approximately unity. Even in this case, however,  the inelastic event rate is expected to be  significant, because of the favorable nuclear structure functions.

In the present paper we discussed the inelastic excitations of $^{125}$Te. For completeness we mention that, in addition to the  isotopes $^{127}$I ,$^{129}$Xe \cite{VerEjSav13} and $^{83}$Kr \cite{VAPSKS15} discussed previously, another possibility  is  the   $^{73}$Ge target, in high energy resolution Ge detectors. 
In short, the present paper, in conjunction with the earlier calculations {\cite{VerEjSav13,VAPSKS15},  indicates that the inelastic scattering opens a new powerful way to search for WIMPs via the SD interaction with the nuclear targets.

{\bf  Acknowledgments}:
This work was partially  supported by ARC Centre of Excellence in Particle Physics at the Terascale, program CE110001104 and ARC DP150103101,  at the University of Adelaide. It was also supported by the  Academy of Finland under the Centre of Excellence Programme 2012-2017 (Nuclear and Accelerator Based Physics Program at JYFL) and the FIDIPRO program. P. Pirinen was supported by a graduate student stipend from the Magnus Ehrnrooth Foundation.
FTA was supported by The National Science Foundation Grant NSF PHY-1307204.

\section{Appendix A: The formalism for the WIMP-nucleus differential event rate}
The expression for the elastic differential event rate is well known, see e.g. \cite{SavVer13},\cite{DivVer13} . For the reader's convenience and to establish our notation we will briefly present the essential ingredients here. We will begin with the more familiar time averaged coherent rate, which   can be cast in the form:
\beq
\left .\frac{d R_0}{ dE_R }\right |_A=\frac{\rho_{\chi}}{m_{\chi}}\,\frac{m_t}{A m_p}\, \left (\frac{\mu_r}{\mu_p} \right )^2\, \sqrt{<\upsilon^2>} \,\frac{1}{Q_0(A)} A^2 \sigma_N^{\mbox{\tiny{coh}}}\left .\left (\frac{d t}{du}\right ) \right |_{\mbox{ \tiny coh}},\, \left .\left (\frac{d t}{d u}\right )\,\right |_{\mbox{\tiny coh}}=\sqrt{\frac{2}{3}}\, a^2 \,F^2(u)  \, \Psi_0(a \sqrt{u})
\label{drdu}
\eeq
with with $\mu_r$ ($\mu_p$) the WIMP-nucleus (nucleon) reduced mass and $A$ is the nuclear mass number. $ m_{\chi}$ is the WIMP mass, $\rho(\chi)$ is the WIMP density in our vicinity, assumed to be 0.3 GeV cm$^{-3}$,  and $m_t$ the mass of the target. $u$ is the recoil energy  $E_R$ in dimensionless units introduced here for convenience, $u=\frac{1}{2}(q b)^2=A m_p b^2 E_R$, with $A$ the nuclear mass number of the target and $b$ the nuclear harmonic oscillator size parameter characterizing the nuclear wave function. It simplifies the expressions for the nuclear form factor and structure functions. In fact:
\begin{equation}
 u=\frac{E_R}{Q_0(A)}~~,~~Q_{0}(A)=[m_pAb^2]^{-1}=40A^{-4/3}\mbox{ MeV}
\label{defineu}
\end{equation}

The factor $\sqrt{2/3}$ in the above expression is  $\upsilon_0/\sqrt{\langle \upsilon ^2\rangle}$ since in Eq. (\ref{drdu}) the WIMP flux is given in units of $\sqrt{\langle \upsilon ^2\rangle}$. In the above expressions  $a=(\sqrt{2} \mu_r b \upsilon_0)^{-1}$, $\upsilon_0$ the velocity of the sun around the center of the galaxy and 
 $F(u)$ is the nuclear form factor. Note that the parameter $a$ depends both on the WIMP mass, the target and the velocity distribution. 

For the spin induced contribution
 one finds for the elastic WIMP-nucleus scattering:
\beq
\left .\frac{d R_0}{ dE_R }\right |_A=\frac{\rho_{\chi}}{m_{\chi}}\,\frac{m_t}{A m_p}\, \left (\frac{\mu_r}{\mu_p} \right )^2\, \sqrt{<\upsilon^2>} \,\frac{1}{Q_0(A)} \, \sigma_A^{\mbox{\tiny{spin}}}\left .\left (\frac{d t}{du}\right ) \right |_{\mbox{ \tiny spin}},\,\left . \left (\frac{d t}{d u}\right )\right |_{\mbox{\tiny spin}}=\sqrt{\frac{2}{3}} \, a^2  \, F_{11}(u)   \,  \Psi_0(a \sqrt{u})
\label{drdus}
\eeq
where $F_{11}$ is the isovector spin response function, i.e. $F_{e\ell}$ for the elastic case and $F_{in}$ for the inelastic one. 
\\We notice that the only difference in the shape of the spectrum between the coherent and the spin comes from nuclear physics. Since the square of the form factor and the spin structure functions have approximately similar shapes, we expect the corresponding differential shapes to be the same, and only the scale to be different.

The function $\Psi_0(x),\,x=\frac{\upsilon_{min}}{\upsilon_0} $  is defined by $\Psi_0(x)=g(\upsilon_{min},\upsilon_E(\alpha))/\upsilon_0$, where $\upsilon_0$ is the velocity of the sun around the center of the galaxy and $\upsilon_E(\alpha)$ the velocity of Earth.
   $g(\upsilon_{min},\upsilon_E(\alpha))$   
 depends on the velocity distribution in the local frame through  the minimum WIMP velocity for a given energy transfer, i.e.
\beq \upsilon_{min}=\sqrt{\frac{A \,m_p \,E_R}{2 \, \mu^2_r}}. \eeq

In the above way of writing the differential event rates we have explicitly separated the three important factors:
\begin{itemize}
\item the kinematics,
\item the nuclear cross section $A^2 \sigma_N$ or $\sigma_A^{\mbox{\tiny{spin}}}$
\item the combined effect of the folding of the velocity distribution and  the form factor or the nuclear structure function.
\end{itemize}
 
For the Maxwell-Boltzmann (M-B ) distribution in the local frame $g$ is defined as
follows:
\begin{align}
&g(\upsilon_{min},\upsilon_E)=\frac{1}{\big
(\sqrt{\pi}\upsilon_0 \big
)^3}\int_{\upsilon_{min}}^{\upsilon_{max}}e^{-(\upsilon^2+2 \vbf .
\vbf_E+\upsilon_E^2)/\upsilon^2_0} \, \upsilon  \,
d\upsilon  \, d \Omega, \nonumber
\\
& \upsilon_{max}=\upsilon_{esc},
\end{align}
where $\vbf_E$ is the velocity of the Earth, including the velocity of the sun
around the galaxy. We have neglected the velocity of the Earth around the sun, since we ignore the time dependence (modulation) of the rates.
The above upper cut-off value in the M-B is usually put in by hand.
 Such a cut-off comes in naturally,
however, in the case of velocity distributions obtained from the halo
WIMP mass density in the Eddington approach \cite{VEROW06}, which,
in certain models, resemble a M-B distribution \cite{JDV09}.


Integrating the above differential rates we obtain the total rate including the time averaged rate 
for each mode  given by:
\beq
R_{\mbox{\tiny coh}}=\frac{\rho_{\chi}}{m_{\chi}} \, \frac{m_t}{A m_p}  \left ( \frac{\mu_r}{\mu_p} \right )^2  \sqrt{<\upsilon^2>} \, A^2 \, \sigma_N^{\mbox{\tiny{coh}}} \, t_{\mbox{\tiny coh}},\quad t_{\mbox{\tiny coh}}=\int_{E_{th}/Q_0(A)}^{(y_{\mbox{\tiny esc}}/a)^2} \, \left .\frac{dt}{du}\right |_{\mbox{\tiny coh}}du,
\label{Eq:Trates}
\eeq
\beq
R_{\mbox{\tiny spin}}=\frac{\rho_{\chi}}{m_{\chi}} \, \frac{m_t}{A m_p} \, \left ( \frac{\mu_r}{\mu_p} \right )^2  \sqrt{<\upsilon^2>} \, \sigma_A^{\mbox{\tiny{spin}}} \, t_{\mbox{\tiny spin}}, \quad t_{\mbox{\tiny spin}}=\int_{E_{th}/Q_0(A)}^{(y_{\mbox{\tiny esc}}/a)^2} \, \left .\frac{dt}{du}\right |_{\mbox{\tiny spin}}du
\label{Eq:Tratec}
\eeq
for each mode (spin and coherent). $E_{th}(A)$ is the energy threshold imposed by the detector.

These expressions contain the following parts: i) The gross properties and kinematics ii) The parameter $t$, which contains the effect of the velocity distribution and the nuclear form factors iii) The WIMP-nuclear cross sections $A^2 \sigma_N$ or $\sigma_A^{\mbox{\tiny{spin}}}$. The latter, contains the nuclear static spin MEs.  From the latter the elementary nucleon cross sections can be obtained, if one mode becomes dominant as already mentioned above.  Using the values for nucleon cross sections, $\sigma_N^{\mbox{\tiny{coh}}}$ in Eq. (\ref{Eq:Trates}) and $ \sigma_A^{\mbox{\tiny{spin}}}$ in Eq. (\ref{Eq:Tratec}), we can obtain the total rates. \\
Conversely, if only one mode is dominant, one can extract from the data the relevant nucleon cross section (coherent, spin isoscalar or spin isovector) or obtain exclusion plots on them. 
\section{Appendix B: Kinematics in the case of  inelastic scattering}

The evaluation of the differential rate for the inelastic transition proceeds in a fashion similar to that of the elastic case discussed above, except:
\begin{enumerate}
\item The transition spin matrix element must be used.
\item The transition spin response function must be used.
For Gamow-Teller like transitions, it does not vanish for zero energy transfer. So it can be normalized to unity, if the static spin value is taken out of the ME.
\item The kinematics is modified.
 The energy-momentum conservation reads:
 \beq
 \frac{-q^2}{2 \mu_r}+\upsilon \xi q -E_x =0, \quad E_x= \mbox {excitation energy}\Leftrightarrow -\frac{m_A}{\mu_r}E_R +\upsilon \xi \sqrt{2 m_A E_R}-E_x =0,
 \eeq
 where $\xi$ is the cosine of the angle between the incident WIMP and the recoiling nucleus. From the above expression we immediately see that  $\xi>0$ as in the case of the elastic scattering. Furthermore the condition $\xi<1$ 
imposes the constraint:
 \beq
  \upsilon>\frac{E_x+\frac{m_A}{\mu_r}E_R}{\sqrt{2 m_A E_R}}
 \eeq
We thus find that for a given energy transfer $E_R$ the minimum allowed WIMP velocity is given by:
\beq
\upsilon_{min}=\frac{E_x+\frac{m_A}{\mu_r}E_R}{\sqrt{2 m_A E_R}}
\eeq
while the maximum and minimum energy transfers are limited by the escape velocity in the WIMP velocity distribution. We find that:
\beq
(E_R)_{min}\leq E_R\leq (E_R)_{max}
\eeq
with
\beq
\left (E_R\right) _{min}=\frac{\mu_r^2}{m_A}\left( \upsilon_{esc}^2-\frac{E_x}{\mu_r}-\sqrt{\upsilon_{esc}^4-2 \upsilon_{esc}^2\frac{E_x}{\mu_r}}\right),\,\left (E_R\right) _{max}=\frac{\mu_r^2}{m_A}\left( \upsilon_{esc}^2-\frac{E_x}{\mu_r}+\sqrt{\upsilon_{esc}^4-2 \upsilon_{esc}^2\frac{E_x}{\mu_r}}\right)
\eeq
In the case of elastic scattering  we recover the familiar formulas:
$$(E_R)_{min}=0,\,(E_R)_{max}=2 \frac{\mu_r^2}{m_A}\upsilon_{esc}^2$$
 From the above expressions it is clear that for a given nucleus and excitation energy only WIMPs with a mass above a certain limit are capable of causing the inelastic transition, i.e. 
\beq
m_{\chi}\geq m_A\left (\frac{1}{2}\upsilon^2_{esc}\frac{m_A}{E_x}-1\right )^{-1}\rightarrow
\eeq
$$  \left (m_{\chi}\right )_{min}=(4.6,\,19,\,34,\,21)\mbox{ GeV for } ^{83}\mbox{Kr},\,^{125}\mbox{Te},\,^{127}\mbox{I},\,^{129}\mbox{Xe  respectively}$$

 We find it simpler to  deal with the phase space in dimensionless units. Noticing that $u=(1/2) q^2 b^2$ and
 \beq
 \delta \left (\frac{-q^2}{2 \mu_r}+\upsilon \xi q -E_x \right )=\delta \left (-\frac{u}{\mu_r b^2}+\upsilon \xi\frac{\sqrt{2 u}}{b}-E_x \right )\Leftrightarrow \frac{b}{\upsilon \sqrt{2u}}\delta \left (\xi-\frac{E_x+u/(\mu_r b^2)}{\upsilon \sqrt{2u}}\right )
 \eeq
 we find:
 \beq
 \int q^2 d \xi dq \delta \left (\frac{-q^2}{2 \mu_r}+\upsilon \xi q -E_x \right )=\frac{1}{b^2 \upsilon} du
 \eeq
 i.e. we recover the same expression as in the case of ground state transitions.\\
 We  now get
 \beq
 y>a \frac{u+u_0}{\sqrt{u}}, y=\frac{\upsilon}{\upsilon_0}
 \eeq
 \beq
 u_0=\mu_r E_x b^2,\quad a=\frac{1}{\sqrt{2}\mu_r \upsilon_0 b},\quad u=\frac{E_R}{Q_0(A)}.
 \eeq
It should be stressed that for transitions to excited states the energy of recoiling nucleus must be above a minimum energy, which depends on the escape velocity, the excitation energy and the mass of the nucleus as well as the WIMP mass. This limits the inelastic scattering only for recoiling  energies  above  the values $\left(E_R\right)_{min}$.
 The minimum and maximum energy that can be transferred is:
 \beq
 u_{min}=\frac{1}{4}\left (\frac{y_{esc}}{a}-\sqrt{\left (\frac{y_{esc}}{a} \right )^2-4 u_0} \right )^2,\,
 u_{max}=\frac{1}{4}\left (\frac{y_{esc}}{a}+\sqrt{\left (\frac{y_{esc}}{a} \right )^2-4 u_0} \right )^2
 \eeq
 \end{enumerate}
 The maximum energy transfers $ u_{max}$  depend on the escape velocity $\upsilon_{exc}=620$km/s. Here we have denoted   $\upsilon_{exc}=y_{esc}\upsilon_0$, where $\upsilon_0$ is the characteristic velocity of the M-B distribution taken to be 220 km/s, i.e. $y_{esc}\approx2.84$. 
The values $u_{min}$, $u_{max}$, $\left(E_R\right)_{min}$ and $\left(E_R\right)_{max}$ relevant for the inelastic scattering of $^{83}$Kr and $^{125}$Te  can be found in our eralier work \cite{VAPSKS15}.

\end{document}